# Two-Dimensional Multi-Pole Solitons

# in Nonlocal Nonlinear Media

Carmel Rotschild,[1] Zhiyong Xu,[2] Oren Cohen,[3] Yaroslav V. Kartashov,[2]

Lluis Torner,[2] and Mordechai Segev[1]

1    *Physics Department and Solid State Institute, Technion, Haifa 32000, Israel*
2    *ICFO-Institut de Ciencies Fotoniques and Universitat Politecnica de Catalunya,*
     *Mediterranean Technology Park, 08860 Castelldefels (Barcelona), Spain*
3    *JILA and Department of Physics, University of Colorado, Boulder, CO, 80309-0440, USA*

We present the experimental observation of scalar multi-pole solitons in highly nonlocal nonlinear media, including dipole-, tri-pole, quadru-pole, and necklace-type solitons, organized as arrays of out-of-phase bright spots. These complex solitons are meta-stable, but with a large parameters range where the instability is weak, enabling their experimental observation.

The formation of solitons in nonlinear medium is one of the most interesting phenomena encountered in nonlinear optics. Solitons can take on complex forms, such as dipole solitons [1,2], multi-hump solitons [3], solitons organized as necklaces [4-8], and even complex beams carrying angular momentum, like rotating propellers [9]. Typically, bright solitons possessing complicated forms necessitate the presence of multiple fields, i.e., are vector (composite) solitons [1-4,6-8]. In fact, in local nonlinear media, the only examples of multi-hump scalar solitons are necklace solitons [4,5]: rings of out-of-phase bright spots holding each other together and arresting the instabilities. The diffraction broadening of such beam is indeed eliminated by the nonlinearity, yet these scalar self-trapped necklace beams still inevitably (slowly) expand, because there is a net outward force exerted on each spot by all other spots comprising the ring [4,5]. Adding angular momentum to the necklace introduces rotation that slows down the expansion but never stops it completely [5]. Thus, the general conclusion is that scalar solitons in homogeneous local nonlinear media cannot form complex states. The picture changes drastically when the nonlinear material response is nonlocal. Nonlocality has profound effects on the complexity of solitons since it allows to overcome repulsion between out-of-phase bright [10-17] or in-phase dark solitons [18] that can form bound states observed in one-dimensional settings [19,20]. In two transverse dimensions, however, the only complex structures thus far observed with scalar solitons were bright vortex-rings [21]. Even though simplest bound states of 2D solitons in nonlocal media was predicted in 80s [11], they still were not observed experimentally.

Here, we present the experimental observation of various types of multi-pole scalar solitons in thermal nonlocal nonlinear medium. We find that multipole solitons in such medium are oscillatory unstable, yet their instability decay rates can be very small under appropriate conditions, giving rise to experimentally accessible metastable complex soliton states.

Our system is described by the evolution equation for the slowly varying light field amplitude $A$ coupled to the steady-state heat transfer equation describing the temperature distribution in the lead glass sample [21]. The light beam is slightly absorbed, and acts as a heat source. Heat diffuses, creating non-uniform temperature distribution, which gives rise to refractive index change proportional to the temperature change. The resulting system of equations in dimensionless form reads as [21]

$$i\frac{\partial q}{\partial \xi} = -\frac{1}{2}\left(\frac{\partial^2 q}{\partial \eta^2} + \frac{\partial^2 q}{\partial \zeta^2}\right) - nq,$$
$$\frac{\partial^2 n}{\partial \eta^2} + \frac{\partial^2 n}{\partial \zeta^2} = -|q|^2. \tag{1}$$

Here $q = (k_0 w_0^2 \alpha\beta / \kappa n_0)^{1/2} A$ is the dimensionless light field amplitude; $n = k_0^2 w_0^2 \delta n / n_0$ is proportional to nonlinear change $\delta n$ in the refractive index $n_0$; $\alpha, \beta, \kappa$ are the optical absorption coefficient, the thermal dependence of the refractive index ($\beta = dn/dT$), and the thermal conductivity coefficient, respectively; the transverse coordinates $\eta, \zeta$ are scaled to the beam width $w_0$, while the longitudinal coordinate $\xi$ is scaled to diffraction length $k_0 w_0^2$. In our lead glass sample $n_0 = 1.8$, the thermal coefficient is $\beta = 14\times 10^{-6}\,\mathrm{K}^{-1}$, the absorption coefficient is $\alpha \approx 0.01\,\mathrm{cm}^{-1}$ and thermal conductivity is $\kappa = 0.7\,\mathrm{W/(mK)}$. Such glass parameters are sufficient to support solitons with widths $\sim 50\,\mu\mathrm{m}$, which give rise to an index change $\delta n \sim 5\times 10^{-5}$ for a total optical power $\sim 1\,\mathrm{W}$. Notice, that system (1) conserves the energy flow $U = \int\int_{-\infty}^{\infty} |q|^2\, d\eta d\zeta$.

We search for soliton solutions of Eq. (1) of the form $q(\eta,\zeta,\xi) = w(\eta,\zeta)\exp(ib\xi)$, where $w(\eta,\zeta)$ is a real function and $b$ is the propagation constant. The soliton intensity vanishes at the boundaries of integration window, while the refractive index $n \to n_b$, where the limiting value $n_b$ is related to the temperature of the sample boundaries, which are kept at fixed and equal temperature. Mathematically, adding the constant background $n_b$ in the refractive index is equivalent to a shift of propagation constant $b$ by the same amount, henceforth we set $n_b = 0$. In this case, the soliton properties are determined solely by $b$ and the width of integration window. We then set the window size $\eta,\zeta \in [-20, 20]$, closely resembling the actual transverse size of our sample. We find numerically a variety of well-localized multi-pole solitons, and test their stability by propagating them numerically in the presence of complex noise. Figure 1 shows illustrative examples of multi-pole solitons, including dipole [Fig. 1(b)], tri-pole [Fig. 1(d)], and quadru-pole [Fig. 1(f)] solitons, as well as necklace solitons [Fig. 1(h)] comprising several bright spots with phase changing by $\pi$ between adjacent spots. In a highly nonlocal nonlinear medium, the refractive index is determined by the intensity distribution in the entire transverse plane, and under proper conditions the nonlocality can lead to an increase of refractive index in the overlap region between out-of-phase solitons even when intensity there is zero, thus giving rise to formation of multipole solitons. Note that the width of the refractive index distribution (the light-induced potential) greatly exceeds the width of an individual light spot. This is a direct indication of the very large range of nonlocality in thermal media. We find that for all types of solitons the energy flow monotonically increases with $b$, which is accompanied by a decrease in the integral soliton width. Similarly, the separation $\delta W$ between the intensity maxima of the multi-pole solitons is also found to decreases with $b$.

Our experiments are carried out in lead-glass samples with a square 2 mm x 2 mm cross-section, which are 84 mm long in the propagation direction. All four transverse boundaries of the sample are thermally connected to a heat sink and maintained at a fixed temperature. In these experiments., we use an 1.8 Watt laser beam at 488 nm wavelength. We launch the dipole soliton by introducing a $\pi$ phase jump across the Gaussian laser beam by inserting a piece of flat glass (of a proper thickness) through one half of the beam cross-section, and imaging it (demagnified) onto the input face of the sample at normal incidence. We launch the tri-pole soliton in a similar fashion, with two parallel pieces of glass, each introducing $\pi$ phase-delay, passing through one-third, and two-thirds of the beam cross-section, respectively. For the quadru-pole soliton, we use two $\pi$ phase-delays, organized perpendicular to one another in the transverse plane, and each passing through one half of the laser beam. Finally, in order to create the sixteen-lobe necklace soliton, we reflect the laser beam off a properly designed phase mask, and subsequently image the beam onto the input face of the sample. We monitor the intensity distribution at the input and output faces by imaging the input and output beams onto a CCD camera. Typical experimental results, with comparisons to the theoretical simulations, are summarized in Fig. 1. The left column of each row shows the input beam in each case. At low power (10 mWatt), the beams linearly diffract for 84mm, after which they broaden significantly (middle columns). At high power (1.8Watt), each beam forms a soliton, which maintains its intensity profile while propagating for 84mm (right colums).

Extensive simulations of the propagation dynamics of perturbed solitons reveal that, in fact, all multi-pole solitons in thermal media are oscillatory unstable. Small perturbations on the input field distribution cause progressively increasing oscillations in the intensities of bright spots comprising the soliton, leading eventually to the destruction of the multi-pole soliton structure. For example, Fig. 2 shows the long-range dynamics of a perturbed (5% complex amplitude noise) dipole soliton, and its transformation into a ground-state soliton, for two values of $b$. The strength of the instability dramatically decreases with decreasing energy flow $U$, so that already at moderate energy levels the solitons survive over large distances (hundreds of diffraction lengths), greatly exceeding the present experimentally feasible sample lengths. We emphasize that we find the necklace solitons also to be meta-stable in our nonlocal thermal media. To our knowledge, these necklaces are the only known case where nonlocality acts to **destabilize** a self-trapped structure (that in this case is not stationary, but is otherwise robust in local nonlinear media [4]), in contrast to the natural tendency of nonlocality to stabilize self-trapped states [22-24].

In conclusion, we have demonstrated experimentally two-dimensional metastable multi-pole solitons in highly nonlocal nonlinear media. The long range of nonlocality enables the formation of a variety of scalar solitons possessing complex structures, varying from dipole solitons, to tri-poles, quadru-poles, to necklaces. Such high nonlocality should be able to support even complex soliton structures carrying angular momentum [25]. This is indeed our next experimental challenge.

This work has been supported by the Israeli Science Foundation, the Generalitat de Catalunya, and by the Government of Spain through the Ramon-y-Cajal program.

# References with titles

# Figure captions

**Figure 1.** (color online) Experimental and theoretical observations of complex scalar solitons in the form of dipole solitons (a,b), tri-pole solitons (c,d), quadro-pole solitons (e,f), and necklace solitons (g,h). The left columns show the input beams, the central columns show the output beams after linear diffraction–broadening for 84 mm propagation, and the right columns show the high-power self-trapped output beams after the same distance.

Figure 2. (color online) Propagation dynamics of slightly perturbed dipole-mode solitons with $b = 3$ (a) and $b = 12$ (b).

Figure1:

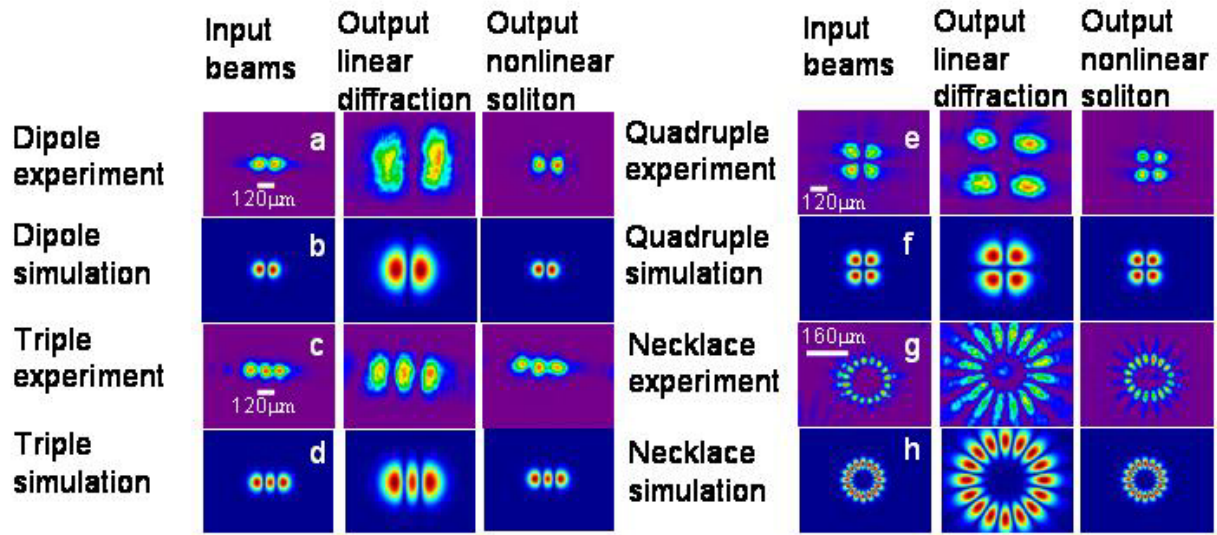

(color online) Experimental and theoretical observations of complex scalar solitons in the form of dipole solitons (a,b), tri-pole solitons (c,d), quadro-pole solitons (e,f), and necklace solitons (g,h). The left columns show the input beams, the central columns show the output beams after linear diffraction–broadening for 84 mm propagation, and the right columns show the high-power self-trapped output beams after the same distance.

Figure 2:

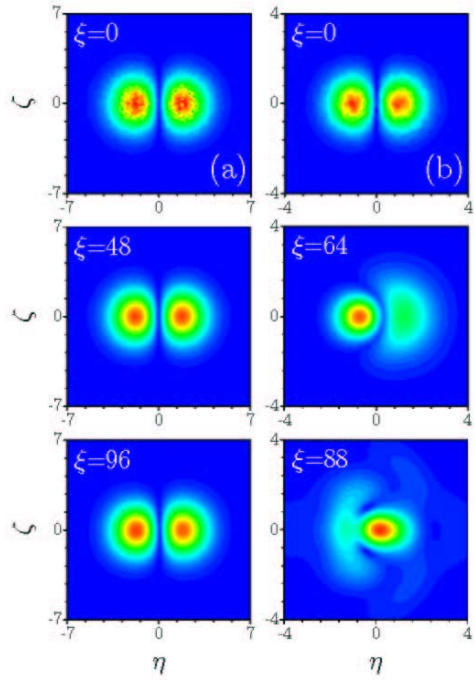

(color online) Propagation dynamics of slightly perturbed dipole-mode solitons with $b = 3$ (a) and $b = 12$ (b).